\documentclass[final,3p,times,twocolumn]{elsarticle}

 \usepackage{graphicx}

\usepackage{amssymb}

\usepackage{url}

\begin{document}

\begin{frontmatter}



\title{Search for $K^-pp$ bound state via $\gamma d \rightarrow K^+ \pi^-X$ reaction at $E_\gamma=1.5-2.4$ GeV}


\author [1] {A.O.Tokiyasu}
\author[2]{M.~Niiyama}
\author[2]{J.D.~Parker}
\author[1]{D.S.~Ahn}
\author[3]{J.K.~Ahn}
\author[1]{S.~Ajimura} 
\author[4]{H.~Akimune}
\author[5]{Y.~Asano}
\author[6]{W.C.~Chang}
\author[1]{J.Y.~Chen}
\author[7]{S.~Dat\'{e}}
\author[1]{H.~Ejiri}
\author[34]{H.~Fujimura}
\author[1,9]{M.~Fujiwara}
\author[10]{S.~Fukui}
\author[1]{S.~Hasegawa}
\author[11]{K.~Hicks}
\author[1]{K.~Horie}
\author[1]{T.~Hotta}
\author[12]{S.H.~Hwang}
\author[12]{K.~Imai}
\author[8]{T.~Ishikawa}
\author[13]{T.~Iwata}
\author[14]{Y.~Kato}
\author[15]{H.~Kawai}
\author[1]{K.~Kino}
\author[1]{H.~Kohri}
\author[1]{Y.~Kon}
\author[7]{N.~Kumagai}
\author[6]{D.L.~Lin}
\author[1]{Y.~Maeda}
\author[17]{S.~Makino}
\author[18]{T.~Matsuda}
\author[19]{T.~Matsumura}
\author[1]{N.~Matsuoka}
\author[1]{T.~Mibe}
\author[8]{M.~Miyabe}
\author[20]{M.~Miyachi}
\author[8]{N.~Muramatsu}
\author[22]{R.~Murayama}
\author[1]{T.~Nakano}
\author[1]{Y.~Nakatsugawa}
\author[22]{M.~Nomachi}
\author[7]{Y.~Ohashi}
\author[7]{H.~Ohkuma}
\author[1]{T.~Ohta}
\author[15]{T.~Ooba}
\author[6]{D.S.~Oshuev}
\author[24]{C.~Rangacharyulu}
\author[1,3]{S.Y.~Ryu}
\author[22]{A.~Sakaguchi}
\author[1]{T.~Sawada}
\author[25]{P.M.~Shagin}
\author[15]{Y.~Shiino}
\author[8]{H.~Shimizu}
\author[32]{E.A.Strokovsky}
\author[22]{Y.~Sugaya}
\author[26]{M.~Sumihama}
\author[33]{J.L.~Tang}
\author[18]{Y.~Toi}
\author[7]{H.~Toyokawa}
\author[2]{T.~Tsunemi}
\author[27]{M.~Uchida}
\author[28]{M.~Ungaro}
\author[29]{A.~Wakai}
\author[6]{C.W.~Wang}
\author[6]{S.C.~Wang}
\author[4]{K.~Yonehara}
\author[1,7]{T.~Yorita}
\author[30]{M.~Yoshimura}
\author[1]{M.~Yosoi}
\author[31]{R.G.T.~Zegers}
\author{(LEPS Collaboration)}

\address[1]{Research Center for Nuclear Physics, Osaka University, Ibaraki, Osaka 567-0047, Japan}
\address[2]{Department of Physics, Kyoto University, Kyoto 606-8502, Japan}
\address[3]{Department of Physics, Pusan National University, Busan 609-735, Republic of Korea}
\address[4]{Department of Physics, Konan University, Kobe, Hyogo 658-8501, Japan}
\address[5]{XFEL Project Head Office, RIKEN 1-1, Koto Sayo Hyogo 679-5148, Japan}
\address[6]{Institute of Physics, Academia Sinica, Taipei 11529, Taiwan}
\address[7]{Japan Synchrotron Radiation Research Institute, Sayo, Hyogo 679-5143, Japan}
\address[8]{Research Center for Electron Photon Science, Tohoku University, Sendai, Miyagi 982-0826, Japan}
\address[9]{Advanced Photon Research Center, Japan Atomic Energy Agency, Kizugawa, Kyoto 619-0215, Japan}
\address[10]{Department of Physics and Astrophysics, Nagoya University, Nagoya, Aichi 464-8602, Japan}
\address[11]{Department of Physics And Astronomy, Ohio University, Athens, Ohio 45701, USA}
\address[12]{Advanced Science Research Center (ASRC), Japan Atomic Energy Agency (JAEA), Tokai, Ibaraki 319-1195, Japan}
\address[13]{Department of Physics, Yamagata University, Yamagata 990-8560, Japan}
\address[14]{ Division of Particle and Astrophysical Sciences,Nagoya University, Furo-cho, Chikusa-ku, Nagoya-shi 464-6802, Japan}
\address[15]{Department of Physics, Chiba University, Chiba 263-8522, Japan}

\address[17]{Wakayama Medical College, Wakayama, Wakayama 641-8509, Japan}
\address[18]{Department of Applied Physics, Miyazaki University, Miyazaki 889-2192, Japan}
\address[19]{Department of Applied Physics, National Defense Academy in Japan, Yokosuka, Kanagawa 239-8686, Japan}
\address[20]{Department of Physics, Tokyo Institute of Technology, Tokyo 152-8551, Japan} 
\address[22]{Department of Physics, Osaka University, Toyonaka, Osaka 560-0043, Japan}
\address[24]{Department of Physics and Engineering Physics, University of Saskatchewan, Saskatoon SK S7N 5E2, Canada}

\address[25]{School of Physics and Astronomy, University of Minnesota, Minneapolis, Minnesota 55455, USA}
\address[26]{Department of Education, Gifu University, Gifu 501-1193, Japan}

\address[27]{Department of Physics, Tokyo Institute of Technology, Tokyo 152-8551, Japan}
\address[29]{Akita Research Institute of Brain and Blood Vessels, Akita 010-0874, Japan}
\address[30]{Institute for Protein Research, Osaka University, Suita, Osaka 565-0871, Japan}
\address[31]{National Superconducting Cyclotron Laboratory, Michigan State University, East Lansing, Michigan 48824, USA}
\address[28]{Department of Physics, University of Connecticut, Storrs, Connecticut 06269-3046, USA}
\address[32]{Joint Institute for Nuclear Research, RU-141980 Dubna, Russia}
\address[33]{Department of Physics, National Chung Cheng University, Taiwan}
\address[34]{School of Medicine, Wakayama Medical University, Wakayama, 641-0011, Japan}

\begin{abstract}
A search for $K^-pp$ bound state (the lightest kaonic nucleus) has been performed using the $\gamma d \rightarrow  K^+ \pi^- \rm{X}$ reaction at E$_\gamma$=1.5-2.4 GeV at LEPS/SPring-8. 
The differential cross section of $K^+ \pi^-$ photo-production off deuterium has been measured for the first time in this energy region, and a bump structure was searched for in the inclusive missing mass spectrum. 
A statistically significant bump structure was not observed in the region from 2.22 to 2.36 GeV/$c^2$, and the upper limits of the differential cross section for the $K^-pp$ bound state production were determined to be 0.1$-$0.7 $\mu$ b (95$\%$ confidence level) for a set of assumed binding energy and width values. 
\end{abstract}


\end{frontmatter}


\section{Introduction}
\label{sec:intro}
Kaonic nuclei provide us with rich information on the sub-threshold $\bar{K}N$ interaction and the nature of $\Lambda(1405)$ in the nuclear medium. 
Since the $\bar{K}N$ interaction is strongly attractive in isospin 0 channel, the existence of  kaonic nuclei is supported theoretically. 
Many experiments have been performed to search for such states using various reactions. 
KEK-PS \mbox{E471$/$E549} group searched for the four-body systems: $K^- ppn$ and $K^- pnn$ using the stopped $K^-$ reaction on a liquid $^{4}$He target \cite{Sato:2007sb, Yim:2010zza}. 
Narrow (the width is less than $40$ MeV) peaks were not observed in the missing mass spectra of $^{4}$He$(K^-_{stopped},p)$X and $^{4}$He$(K^-_{stopped},n)$X, and the group concluded that the upper limits of the formation probability are below a few percent per stopped $K^-$ events. 
Osaka group studied the $\bar{K}$-nucleus interaction using in-flight $(K^-, N)$ reactions on $^{12}$C and $^{16}$O targets \cite{Kishimoto:2005, Kishimoto:2007}. 
The authors observed an enhancement of the yield in the $K^-$ bound region in the inclusive spectra. 
The shapes of the obtained spectra were compared with theoretical calculations, and the derived $\bar{K}$ potential appeared as rather deep ($160-190$ MeV). 
However another theoretical calculation was also able to explain the spectra with a shallow (60 MeV)  potential \cite{Magas:2010}, and more intensive experimental investigations are necessary to improve the precision of the shape analysis. 

The lightest system, $\bar{K}NN$ is a fascinating system to investigate the sub-threshold $\bar{K}N$ interaction more precisely. 
In particular, the bound state of $K^-$ and two protons ($K^-pp$ bound state) has been actively studied because it has the largest number of $\bar{K}N$ pairs with I=0 and is estimated to be the strongest binding system among the three-body systems. 
The structure and the production mechanism of $K^-pp$ bound state have been investigated using various theoretical approaches \cite{Yamazaki:2007, Ikeda:2007, Shevchenko:2006xy, Shevchenko:2007zz, Ikeda:2010, Uchino:2011, Dote:2009}. 
The binding energy (B.E.) and the width ($\Gamma$) were predicted to be $9-95$ MeV and $34-110$ MeV, respectively. 
The predicted values are in considerable disagreement depending on the $\bar{K}N$ interaction models and the calculation methods. 
On the other hand, $K^-pp$ bound state has been searched for experimentally, and there are two groups who have detected the possible candidates. 
The first measurement was reported by \mbox{FINUDA} group at \mbox{DA$\Phi$NE}\cite {ref:FINUDA}. 
They investigated the stopped $K^-$ reaction on five kinds of targets ($^{6}$Li, $^{7}$Li, $^{12}$C, $^{27}$Al and $^{51}$V) and observed a bump structure in the invariant mass spectrum of $\Lambda$ and proton which were emitted back-to-back from the targets. 
The B.E. and $\Gamma$ were determined as $115^{+6}_{-5}(stat)^{+3}_{-4}(syst)$ MeV and $67^{+14}_{-11}(stat)^{+2}_{-3}(syst)$ MeV, respectively. 
However there are some theoretical interpretations that the observed peak can be explained by the two-nucleon absorption with the final state interaction of outgoing particles \cite{ref:Magas}. 
\mbox{DISTO} group at \mbox{SATURNE} re-analyzed the dataset of the exclusive $pp\rightarrow pK^+\Lambda$ reaction and observed a bump structure in the missing mass spectrum of $K^+$ \cite{ref:DISTO}. 
The B.E. and $\Gamma$ were determined as $103\pm 3(stat)\pm 5(syst)$ MeV and $118\pm 8(stat)\pm 10(syst)$ MeV, respectively. 
The measured values of B.E. and $\Gamma$ are different between the two groups, and they are inconsistent with any of the existing theoretical predictions. 
Thus, the existence of $K^-pp$ bound state has not been established yet, and new experiments using different reaction could help to resolve the controversial situation.

In this letter, we report on the search results using the $\gamma {\rm d} \rightarrow K^+ \pi^- {\rm X}$ reaction in the photon energy region of $E_\gamma=1.5-2.4$ GeV. 
By using a deuteron target, we can reduce the ambiguity from nuclear effects such as Fermi motions. 
$K^-pp$ bound state is expected to have a spatially broad wave function. 
Therefore, the production cross section is expected to be enlarged with small momentum transfer kinematics. 
In order to investigate the small $|t|$ region, a $\pi^-$ is detected at forward angles in addition to a $K^+$. 
In the photon induced reactions, $\overline{K}$ can be exchanged in the $t$-channel, while it is prohibited in pion or kaon induced reactions because the vertex with three pseudo scalar particle lines is prohibited.
From the beam asymmetry measurements of hyperon photo-productions, a substantial contribution of $t$-channel $K^-$ exchange was confirmed at the very forward regions \cite{ref:LEPS, ref:Hicks}. 
The $t$-channel $K^-$ exchange can be treated as a virtual $K^-$ beam which is close to on-shell by detecting $K^+$ at forward angle. 
From this point of view, we searched for $K^-pp$ bound state via virtual d$(K^-, \pi^-)$X reaction with the four momentum transfer, $|t|$, ranging from 0.2 to 1.4 (GeV/$c$)$^2$. 
A bump structure was searched for in the inclusive missing mass spectrum of d$(\gamma,K^+\pi^-)$X ($MM_{\rm d}(K^+\pi^-)$). 
The study of the the inclusive spectrum allowed us to search for $K^-pp$ bound state without selecting the decay mode, whereas considerable contributions from quasi-free processes arise as the background in the search region. 
These background processes are also discussed in this letter.

\section{The LEPS experiment and Analysis}
\label{sec:exp}
The experiment was performed at LEPS/SPring-8 in 2002/2003 and 2006/2007. 
The experimental conditions and data quality were similar throughout these two data taking periods. 
Therefore two data sets were summed up for the analysis. 
Linearly polarized photons with energy ranging from 1.5 to 2.4 GeV were produced by the backward Compton scattering. 
The energy of each photon was measured by detecting scattered electron with a tagging counter. The photon energy resolution is estimated to be approximately 12 MeV. 
More details about the status of the photon beam are given in \cite{Muramatsu:2012}.

Liquid deuterium (LD$_2$) of 16 cm effective length was used as the target, and 7.6 $\times10^{12}$ tagged photons were incident on the target in total. 
The charged particles produced from the target were detected with the LEPS spectrometer at forward angle in the laboratory system. 
The LEPS spectrometer consists of a start counter (SC), a silica-aerogel $\check{\mathrm{C}}$erenkov counter (AC), a silicon vertex detector (SVTX), drift chambers (DCs) for tracking, a dipole magnet with field strength of 0.7 [T] and a time-of-flight (TOF) scintillator wall.
AC has the reflection index 1.03 and was used for $e^+ e^-$ veto at the trigger level. 
The momentum threshold of AC is 2.0 GeV/$c$ for kaons and 0.57 GeV/$c$ for pions. 
The momenta of the particles were analyzed by tracking information, and particle species were identified using TOF information. 
The momentum resolution is 6 MeV/$c$ at a momentum of 1 GeV/$c$. 
More details about experimental setup are given in \cite{ref:LEPS}. 

For the present analysis, events with $K^+$ and $\pi^-$ tracks were selected with mass values required to be within 3 $\sigma$, where $\sigma$ is the mass resolution depending on the momentum. 
Events for which a $\pi^+$ was misidentified as a $K^+$ were rejected by requiring that the missing mass of p($\gamma, \pi^+\pi^-$)X was above 0.97 GeV/$c^2$, where the $\pi^+$ mass was used for the $K^+$ mass. 
In order to reduce the systematic uncertainty arising from the acceptance correction, the analysis was performed within the following kinematical region: $\cos\theta_{K^+}^{lab}>0.95$, $\cos\theta_{\pi^-}^{lab}>0.95$, 0.25 GeV/$c < p_{K^+}^{lab} <$ 2.0 GeV/$c$, and 0.25 GeV/$c <p_{\pi^-}^{lab} < $0.6 GeV/$c$. 
Here $\theta$ and $p$ denotes the scattering angle and momentum, respectively, and the subscript $"lab"$ indicates the measurement in the laboratory system. 
The vertex points of the $K^+$ and $\pi^-$ tracks were required to be inside the target.
The vertex resolution along the beam axis was approximately 2 mm, and the contamination events from the SC or AC were well-separated from the events from the $\rm{LD}_2$ target.
In addition, the distance of closest approach between the two tracks was required to be less than 4 mm. 
These vertex constraints reduced the contribution of hyperon decay events of which vertex points were outside the target or had large closest distances. 
By comparing the yields of $K^-pp$ bound state and $\Lambda$ generated with a GEANT-based Monte Carlo simulation, the signal to noise ratio was estimated to be improved by a factor of 2. 
Finally, for the events in which three tracks were detected ($K^+$, $\pi^-$, and $p$), the invariant masses of p and $\pi^-$ ($M(p\pi^-)$) were calculated, and events in the range of 1.05 GeV$/c^2< M(p\pi^-) <$ 1.12 GeV$/c^2$ were rejected because they arise from the quasi-free $\Lambda$ production process. 
The event loss due to this cut is small ($\sim 4\%$), and there is little possibility to distort the shape of the spectrum of $MM_{\rm d}(K^+\pi^-)$. 

\section{Results and Discussion}
\label{sec:ana}
\begin{figure}[ht]
 \begin{center}
 \includegraphics*[width=0.45\textwidth]{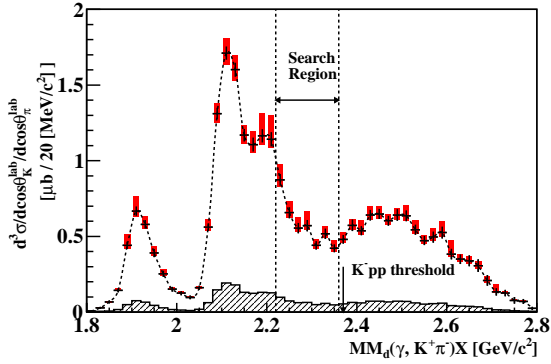}
 \caption{ (color online)(a):  
 Differential cross section of d$(\gamma, K^+\pi^-)$X, $d\sigma/d\cos\theta_{K^+}^{lab}/d\cos_{\pi^-}^{lab}$. Here, bin width is 20 MeV/$c^2$. 
The error bars shows the statistical error, and the red boxes show the systematic error. 
The discrepancy between the two datasets is shown as the hatched histogram.
} 
\label{fig:cs_tot}
 \end{center}
\end{figure}

Figure \ref{fig:cs_tot} shows the differential cross section spectrum of $MM_{\rm d}(K^+\pi^-)$ ($d\sigma/d\cos\theta_{K^+}^{lab}/d\cos_{\pi^-}^{lab})$ after applying the cuts described above. 
The bin width of the spectrum is 20 MeV$/c^2$.
The search region (2.22 GeV$/c^2$-2.36 GeV$/c^2$) and $K^-pp$ mass threshold (2.37 GeV$/c^2$) are also indicated in the figure. 
The differential cross section was obtained by applying an acceptance correction to each track. 
The acceptances were calculated with the GEANT-based Monte Carlo simulation for particles with given one momentum, cos$\theta$, and vertex point \cite{ref:LEPS}. 
The error bars represent the statistical uncertainty, and the boxes represent the systematic uncertainties. 
The systematic uncertainty is dominated by the error in the acceptance correction, which was estimated to be $5.7\%$ in the search region. 
The systematic uncertainty of photon transmission from the collision point to the experimental hutch was estimated to be 3\%.
The misidentified $K^+$ and $\pi^-$ events are distributed mainly in the region below 1.7 GeV/$c^2$ in the $MM_{\rm d}(K^+\pi^-)$ spectrum, and are negligible in the region where the bump structure is sought. 
The fluctuation of the target density was assumed to be negligibly small. 
The systematic uncertainty arising from the above sources was $6.4\%$ in total in the search region. 
An additional systematic uncertainty which is likely due to the normalization of photon numbers  between the two data sets is estimated to be $12\%$ (R.M.S.) using quasi-free $\Lambda$ and $\Sigma$ production. 
This systematic uncertainty is shown as a hatched histogram in Fig.\ref{fig:cs_tot}, obtained by multiplying the spectrum by the constant ratio of 12\%. 

As shown in Fig.\ref{fig:cs_tot}, there are three peaks around 1.9 GeV/$c^2$, 2.1 GeV/$c^2$ and 2.2 GeV/$c^2$ in the $MM_{\rm d}(K^+\pi^-)$ spectrum. 
These peaks correspond to the $\gamma n \rightarrow K^+ \Sigma^-$ followed by $\Sigma^- \rightarrow \pi^- + n$, $\gamma n \rightarrow K^+ \pi^- \Lambda$ and $\gamma p/n \rightarrow K^+ \pi^- \Sigma^{+/0}$ processes. 
The differential cross section of each process was determined as $\sim$3 $\mu$b, $\sim$7 $\mu$b and $\sim$4 $\mu$b, respectively, for the kinematical region used in this analysis. 
If the $K^-pp$ bound state exists and has a large production rate, a bump structure should be observed as the signal in the range of 2.22 GeV/c$^2$ to 2.36 GeV/c$^2$ in the $MM_{\rm d}(K^+\pi^-)$ spectrum. 
Then bump structures corresponding to the production of the $K^-pp$ bound state were searched for with the the Log-likelihood ratio method. 
In this method, the $MM_{\rm d}(K^+\pi^-)$ spectrum was fitted under two hypotheses : background processes only, and background processes and signal process. 
The yield of each background process was considered as a free parameter for the fitting. 
The Log-likelihood value was obtained by fitting the signal and background spectra to the experimental data where the yield of the signal was increased from 0 to a certain value. 
Then the improvements of Log-likelihood from background only hypothesis ($-2 \Delta \ln L$) were tested in the search region. 
It is worthwhile to note that the raw spectrum was used for the fitting because acceptance-corrected spectrum has considerable systematic uncertainties and deteriorates the quality of the fitting. 
Four processes were used for the background: $\gamma n \rightarrow \Lambda K^+ \pi^-$, $\gamma p \rightarrow \Sigma^+ K^+ \pi^-$, $\gamma n \rightarrow \Lambda K^+ \pi^- \pi^0$ and $\gamma p \rightarrow \Sigma(1385)^+ K^+ \pi^-$. 
The shapes of the spectra were generated with the GEANT-based Monte Carlo simulation, where the Paris-potential model was used to describe the momentum distribution of the nucleons inside the deuteron \cite{ref:Paris}. 
In addition, a constant offset was adopted in order to consider the contribution of remaining processes such as hyperon decay. 
\begin{figure}[ht]
 \begin{center}
 \includegraphics*[width=0.5\textwidth]{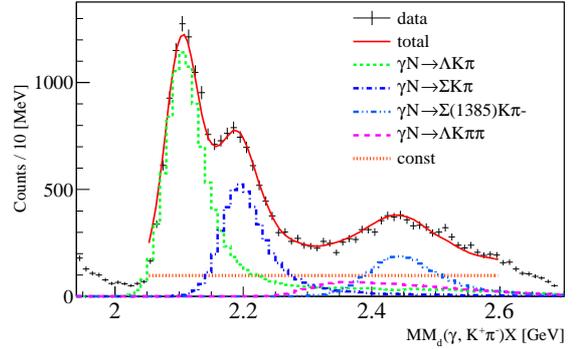}
 \caption{
(color online) 
The spectrum fit result for the determination of the upper limit of cross section.
 The color and style of line for each corresponding process are shown in the figure.
}
 \label{fig:init_fit}
 \end{center}
\end{figure}
Figure \ref{fig:init_fit} shows the fit result with only background processes. 
$\chi^2$/ndf of the fit result is 3.5 in the range from 2.05 GeV/$c^2$ to 2.6 GeV/$c^2$, and approximately 1 in the range from 2.22 GeV/$c^2$ to 2.36 GeV/$c^2$. 
The tests were performed for signals with $\Gamma$ = 20, 60 and 100 MeV, and 15 B.E. values ranging from 10 to 150 MeV. 
The signal shape was assumed to be a Breit Wigner distribution with the fixed B.E. and $\Gamma$ , and was generated with the GEANT-based Monte Carlo simulation. 
As a result of tests, significant improvements of Log-likelihood were not observed under any condition in the search region. 

\begin{figure}[ht]
 \begin{center}
 \includegraphics*[width=0.5\textwidth]{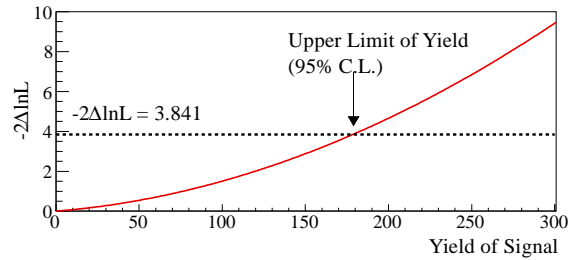}
 \caption{
(color online) Typical $-2\Delta lnL$ as a function of the signal yield. 
 The B.E. and $\Gamma$ is assumed to be 100 MeV and 60 MeV, respectively. 
}
 \label{fig:typ_ex}
 \end{center}
\end{figure}

In order to quantify the search results, the upper limits of the differential cross section were determined. 
The signal yield which gave $-2 \Delta \ln L = 3.84$ was used to give the upper limit of the yield at the 95\% confidence level.
In Fig.\ref{fig:typ_ex}, $-2 \Delta \ln L$ values are shown as a function of the signal yield for B.E.=100 MeV and $\Gamma=$60 MeV as a typical example.
The crossing point at $-2 \Delta \ln L = 3.84$ is indicated by an arrow.

\begin{figure}[h]
 \begin{center}
 \includegraphics*[width=0.5\textwidth]{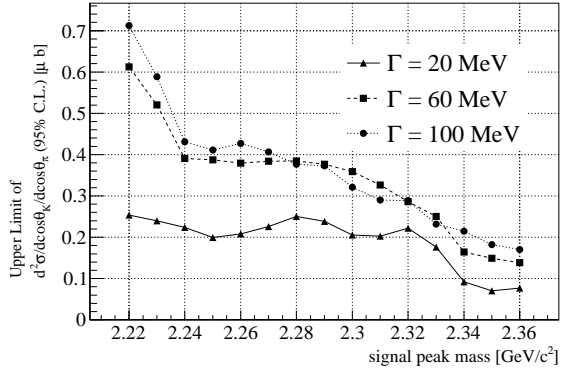}
 \caption{ 
 The upper limit of the differential cross section of $K^-pp$ bound state production as a function of assumed signal peak mass. 
 The solid, broken and dotted lines are the results of $\Gamma=$20 MeV, 60 MeV, and 100 MeV, respectively.
  }
 \label{fig:uplim}
 \end{center}
\end{figure}

Thus, the upper limits of the yield were determined for signals with $\Gamma$ = 20, 60 and 100 MeV, and 15 B.E. values ranging from 10 to 150 MeV. 
The obtained yields were converted to the differential cross section by dividing them by the acceptance of the signals, efficiencies and integrated luminosities. 
The acceptance was determined with the GEANT-based Monte Carlo simulation under the assumption that d($\gamma, K^+\pi^-$)$K^-pp$ reaction occurs isotropically in the center-of-mass system. 
Figure \ref{fig:uplim} shows the upper limits of the differential cross section of $K^-pp$ bound state production for various $\Gamma$ values as a function of the assumed signal peak mass. 

The upper limits of the differential cross section of $K^-pp$ bound state production were determined to be (0.07 $-$ 0.2), (0.1 $-$ 0.6), (0.2 $-$ 0.7) $\mu$b for $\Gamma=20, 60, 100$ MeV, respectively at the 95$\%$ confidence level. 
These values correspond to $0.5\%-5\%$ of the differential cross section of the typical hadron production processes such as $\gamma n \rightarrow K^+ \pi^- \Lambda$ or $\gamma p/n \rightarrow K^+ \pi^- \Sigma^{+/0}$. 
We can compare the obtained results to those of the KEK-PS E471/E549 group.
They concluded that the formation probabilities of the four-body kaonic nuclei are less than a few percent per stopped kaon. 
Since $K^-$ absorbed in nuclei form hyperons, their results equivalently state that the formation probabilities of kaonic nuclei are less than a few percent of the typical hyperon production cross section. 
The obtained search results are comparable with the KEK-PS E471/E549 results although the Kaonic nuclei production mechanisms are expected to be different between the two reactions.

\begin{figure}[ht]
 \begin{center}
  \includegraphics*[width=0.5\textwidth]{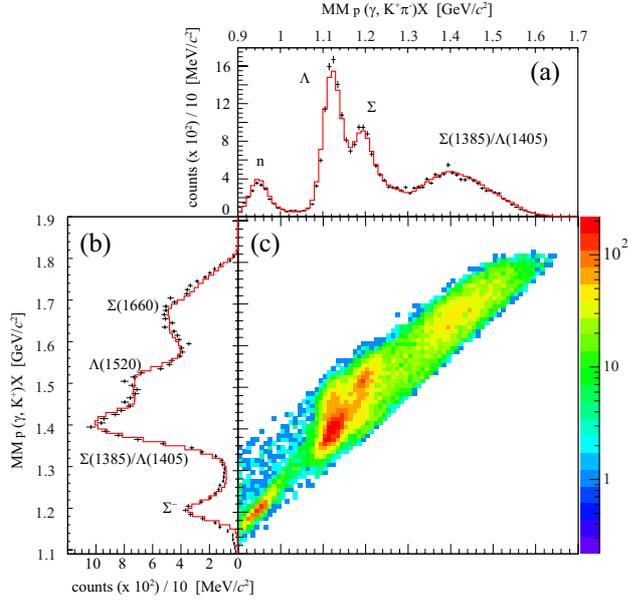}
  \caption{
  (color online)
  Simultaneous fit result of (a) $MM_{\rm p}(K^+\pi^-)$ and (b)$MM_{\rm p}(K^+)$. 
 The process used for fitting is listed in Tab.\ref{tab:proc}. 
 The points with error bar denotes the observed histogram, and 
 red histogram denotes the summation of contribution from each process after the fit.
}
 \label{fig:bg_fit}
 \end{center}
\end{figure}

Though bump structures were not observed, there were several thousand events in the search region. 
In order to investigate the background precisely, the $MM_{\rm p}(K^+\pi^-)$ spectrum and $MM_{\rm p}(K^+)$ spectrum were fitted simultaneously. 
The subscript $p$ means that the missing mass was calculated assuming a proton at rest as the target.
The processes considered for the fitting are listed in Tab.\ref{tab:proc}. 
The contribution of $K^{*0}$ production is negligibly small under the selected kinematic conditions and was ignored. 
PDG values were used for the mean, width, and branching ratio of the hyperon resonances \cite{ref:PDG}, and all the processes were generated isotropically in the center of mass system. 
The mass and width of $\Sigma(1660)$ were assumed to be 1.66 GeV/$c^2$ and 0.1 GeV, respectively, and the branching ratios of the $\Lambda \pi^-$ and $\Sigma \pi^-$ decay modes were considered as free parameters. 
The fit result is shown in Fig.\ref{fig:bg_fit}. 
The experimental data is shown as points with the error bars, and the fit results are shown as a red histogram. 
The total $\chi^2$/ndf is 1.3. 
The fit result indicates that the main contribution to the $MM_{\rm d}(K^+\pi^-)$ spectrum in the search region comes from the $\gamma p \rightarrow K^+ \Lambda(1520)$ process. 
Its fraction of the observed yield is approximately 20\%. 
The non-resonant $\Lambda/\Sigma \pi K^+\pi^-$ production also contributes about 20\% to the signal region. 
The upper limit of production probability of $K^-pp$ bound state was determined to be less than $5\%$ of typical hadron processes, and it was found to be difficult to separated the $K^-pp$ bound state signal from the background processes in the inclusive measurement. 
For the further study, it is necessary to detect the decay products of $K^-pp$ bound state using counters surrounding the target. 
$K^-pp$ bound state is expected to have non-mesonic decay modes such as $ K^-pp \rightarrow \Lambda p$ or $ K^-pp \rightarrow \Sigma N$, and detecting the proton or $\Lambda$ which has large transverse momentum is essential to increase the signal to noise ratio. 

\begin{table}
\begin{center}
\caption{Quasi-Free processes}
\label{tab:proc}
\begin{tabular}{ll}
\hline\hline
proton target & neutron target\\
\hline
$\gamma$ + p $\rightarrow$  $\Lambda$	       $K^+$             &  $\gamma$ + n $\rightarrow$   $\Sigma^-{}$       $K^+$ \\
$\gamma$ + p $\rightarrow$ 	$\Sigma^{0}$	   $K^+$             &  $\gamma$ + n $\rightarrow$   $\Lambda$          $K^+$     $\pi^-$ \\
$\gamma$ + p $\rightarrow$ 	$\Lambda(1405)$   $K^+$             &  $\gamma$ + n $\rightarrow$	 $\Sigma(1385)^-$   $K^+$ \\
$\gamma$ + p $\rightarrow$ 	$\Sigma(1385)^{0}$ $K^+$             &  $\gamma$ + n $\rightarrow$	 $\Sigma(1660)^-$   $K^+$ \\                                                 
$\gamma$ + p $\rightarrow$ 	$\Sigma^{+}$	   $K^+$  $\pi^-$    &  $\gamma$ + n $\rightarrow$   $\Lambda$  $\pi^0$   $K^+$  $\pi^- $ \\
$\gamma$ + p $\rightarrow$	$\Lambda(1520)$	   $K^+$             &   \\
$\gamma$ + p $\rightarrow$	$\Sigma^{0}$ $\pi^+$ $K^+$  $\pi^-$    &   \\
\hline
\end{tabular}
\end{center}
\end{table}

The production cross section of $K^-pp$ bound state is assumed to be dependent on the kinematic condition, especially momentum transfer of residual system. 
Although the production mechanism of $K^-pp$ bound state is poorly understood, if $K^-pp$ bound state was produced via the sticking process of virtual $K^-$ or intermediate resonance states, the kinematic condition of small momentum transfer is efficient to enlarge the production cross section. 
Thus, except for the search under the cut condition described above, a search for a bump structure was done in the different kinematical conditions as follows:
(a) $|t|<0.3$ (GeV/$c$)$^2$, (b) $p_X < 0.8$ GeV/$c$. 
Here, $p_x$ is the momentum transfer in the virtual d$(K^-, \pi^-)X$ reaction. 
In addition, $K^-pp$ bound state was searched for in a different reaction channel : d$(\gamma,K^{*0})$X reaction. 
$K^{*0}$ production events were selected by loosing the cut condition of cos$\theta > 0.95$ and requiring invariant mass of a $K^+$ and a $\pi^-$ to be the $K^{*0}$ mass (0.842 GeV/$c^2$ $< M(K^+\pi^-) <$ 0.942 GeV/$c^2$ ). 
If $K^{*0}$ is produced in $t$-channel, $\overline{K}$ and $\kappa$ exchange is clearly selected \cite{ref:Hwang}. 
A bump structure was searched for in the $MM_d(K^+\pi^-)$ spectrum, but bump structures were not observed under the conditions (a), (b) and (c). 

\section{Conclusion}
\label{sec:concl}
We searched for $K^-pp$ bound state using the $\gamma {\rm d} \rightarrow K^+ \pi^- {\rm X}$ reaction at $E_\gamma=1.5-2.4$ GeV. 
A $K^+$ and a $\pi^-$ were detected at forward angles in order to select the region of small $|t|$. 
The differential cross section of $K^+ \pi^-$ photo-production off deuterium was measured for the first time in this energy region, and a bump structure $K^-pp$ bound state was searched for in the inclusive $MM_{\rm d}(K^+\pi^-)$ spectrum. 
No bump structure was observed in the mass range from 2.22 to 2.36 GeV/$c^2$, and the upper limits of the  differential cross section of $K^-pp$ bound state production were obtained for different B.E. and $\Gamma$. 
The values are (0.07 $-$ 0.2), (0.1 $-$ 0.6) and (0.2 $-$ 0.7) $\mu$b at 95$\%$ confidence level for $\Gamma =$ 20, 60, 100 MeV, respectively.
These values correspond to approximately $0.5-5\%$ of the cross section of typical hadron photo-production.
In order to reveal the origin of the background, the $MM_{\rm p}(K^+ \pi^-)$ spectrum and the $MM_{\rm p}(K^+)$ spectrum were fitted simultaneously with 15 background processes. 
The $\gamma p \rightarrow K^+ \Lambda(1520)$ process and $\gamma N \rightarrow K^+ \pi^- \pi \Lambda/\Sigma $ process were found to be the main contribution in the search region. 

\section*{Acknowledgments}

The authors thank the SPring-8 staff for their contributions in the operation of the LEPS experiment. 
This research was supported in part by the
Ministry of Education, Science, Sports and Culture of Japan, by the GCOE program in Kyoto University, by the
National Science Council of the Republic of China (Taiwan), by the National Research Foundation of Korea (2012R1A2A1A01011926) and by the National Science Foundation
(NSF Award PHY-0244999). 
Author J.D. Parker was supported by a Postdoctoral Fellowship for Foreign Researchers from the Japan Society for the Promotion of Science.

\bibliographystyle{plain}
\bibliography{<your-bib-database>}

\end{document}